\documentclass{article}

\usepackage{spconf}
\usepackage{url}
\usepackage{booktabs}
\usepackage{multirow}
\usepackage{tabularx}

\usepackage{afterpage}
\usepackage{float}

\usepackage{hyperref}
\usepackage{tablefootnote}

\usepackage{amssymb}
\usepackage{siunitx}

\usepackage[english]{babel}

\usepackage{amsmath}
\usepackage{graphicx}

\title{Online Speaker Diarization of Meetings \\ Guided by Speech Separation}
\name{Elio Gruttadauria$^{1}$,
    \quad Mathieu Fontaine$^{1}$, 
    \quad Slim Essid$^{1}$\thanks{This work was supported by the Audible project, funded by French BPI and partly supported by  ANR Project SAROUMANE (ANR-22-CE23-0011). Also, it was performed using HPC resources from GENCI-IDRIS.}}
  
\address{$^{1}$LTCI, T\'el\'ecom Paris, Institut Polytechnique de Paris, France}

\ninept
\begin{document}
\maketitle

\begin{abstract}
Overlapped speech is notoriously problematic for speaker diarization systems. Consequently, the use of speech separation has recently been proposed to improve their performance.
Although promising, speech separation models struggle with realistic data because they are trained on simulated mixtures with a fixed number of speakers.
In this work, we introduce a new speech separation-guided diarization scheme suitable for the online speaker diarization of long meeting recordings with a variable number of speakers, as present in the AMI corpus.
We envisage ConvTasNet and DPRNN as alternatives for the separation networks, with two or three output sources.
To obtain the speaker diarization result, voice activity detection is applied on each estimated source.
The final model is fine-tuned end-to-end, after first adapting the separation to real data using AMI.
The system operates on short segments, and inference is performed by stitching the local predictions using speaker embeddings and incremental clustering. 
The results show that our system improves the state-of-the-art on the AMI headset mix, using no oracle information and under full evaluation (no collar and including overlapped speech).
Finally, we show the strength of our system particularly on overlapped speech sections.
\end{abstract}


\section{Introduction}
Speaker diarization (SD) aims at answering the question ``who spoke when?" by segmenting a recording into speaker-homogeneous regions \cite{park_review_2021}.
\par
Speaker diarization has traditionally been framed as a clustering problem, with systems consisting of a cascade of several steps \cite{sahidullah_speed_2019, landini_analysis_2021}, each individually optimized.
As diarization systems become more effective, the inability of clustering-based systems to model overlapped speech directly becomes a non-negligible limiting factor.
Indeed, up to 20\% of total conversational speech time can be categorized as overlapping speech \cite{watanabe_chime-6_2020}, which naturally calls for a change of paradigm.
End-to-end neural diarization (EEND) models \cite{fujita_end--end_2020, horiguchi_encoder-decoder_2022} reframe the diarization task as a multi-label classification problem. 
By doing this, the EEND framework inherently considers the issue of overlapping speech.
Other examples of non-clustering based systems are target-speaker VAD (TS-VAD) \cite{medennikov_target-speaker_2020} and region proposal networks (RPNs) \cite{huang_speaker_2020}.
Although EEND-based systems have shown state-of-the-art performance over the clustering paradigm, the best models rely on the self-attention mechanism \cite{vaswani_attention_2023} and tend to require a lot of data to be trained properly. 
In this context, speech separation models (SSep) show potential for better handling overlapped speech, 
while being computationally more efficient \cite{tzinis_sudo_2020}.
Currently, the novel speech separation guided diarization (SSGD) paradigm \cite{fang_deep_2021, niu_separation_2021} is still limited because of the inability of SSep models to behave well on realistic data: the better performance on overlapped speech sections is counteracted with worse performance on the remainder of the audio.
Additionally, no work has been done yet to deal with multiple speakers (\textit{i.e.}, more than 2 speakers), making the SSGD paradigm not ready yet for general settings, even less for online speaker diarization.
\par
Online SD systems make predictions at each time step with information available only up until that point (or slightly in the future).
Only a few models are online by nature \cite{zhang_fully_2019}, but offline systems may sometimes be adapted to operate online.
The work from Kinoshita et al. \cite{kinoshita_integrating_2021, kinoshita_advances_2021} introduces an adaptation of the EEND model to handle long recordings.
Coria et al. \cite{coria_overlap-aware_2021} used the same technique to adapt the EEND framework to real-time processing.
In their proposal, predictions are made locally on short overlapping windows, and incremental clustering is used to solve the permutation problem.
\par
With this work, we introduce a novel speaker diarization system architecture that expands the SSGD paradigm to accommodate meeting recordings (with more than 2 speakers), and we study its performance in the online diarization setting, focusing on single-microphone scenarios.
To the best of our knowledge, this is the first work using SSep for diarization outside the conversational telephone speech (CTS) domain where only 2 speakers are present in the entire recording.
As will be discussed in section \ref{sec:SS_different_outputs}, separation models struggle when the number of speakers active during the testing phase differs from that considered during the training of the separation network.
Nevertheless, our solution is suitable for an arbitrary number of speakers. 
Notably, we are able to improve the state-of-the-art performance on AMI headset mix in the online setting using no oracle information.
Our system can also estimate sources for each speaker in addition to the diarization result. 
Finally, we also show the superiority of our method on the overlapped speech sections in particular.
The code to reproduce the results of this work is freely available\footnote{\href{https://github.com/egruttadauria98/SSpaVAlDo}{egruttadauria98/SSpaVAlDo}}.

\section{Related work}
Fang et al. \cite{fang_deep_2021} introduced the speech separation guided diarization (SSGD) approach, refining the work from \cite{niu_separation_2021}.
Their system employs dynamic selection between conventional clustering-based diarization which is effective for single-speaker segments and SSGD which excels in handling overlapped speech.
However, they note occasional SSGD instability and SSep model failures, resulting in speaker confusion and false alarms due to channel leakage and artifacts in the estimated sources.
In the context of SSep models, leakage is defined as the presence of one or more other speakers in an estimated source.
In \cite{morrone_low-latency_2023}, a leakage removal algorithm is proposed, based on the SI-SDR metric \cite{roux_sdr_2018}.
Recent advancements \cite{morrone_end--end_2023} show that fine-tuning the model is an effective approach to mitigate the source leakage problem. 
Notably, adapting the VAD to the estimated sources reduces false alarms, but the best results arise from jointly fine-tuning the separation model and VAD in an end-to-end manner. 
\par
The previous work on SSGD research focuses on the conversational telephone speech (CTS) domain, involving at most two speakers.
This limitation simplifies the SSGD problem enabling a single-pass inference to be performed.
Alternatively, for longer recordings or SSep models with a restricted receptive field, the system can operate on short overlapping windows whose predictions are then stitched together using the correlation between overlapping sections of consecutive windows.
This approach is known as Continuous Speech Separation (CSS) \cite{chen_continuous_2020}.
When working with meeting conversations, where more than two speakers are present, the CSS approach is no longer feasible, as it implies that each local prediction must have as many outputs as the total number of speakers.
In fact, as discussed in Section \ref{sec:SS_different_outputs}, SSep models like ConvTasNet \cite{luo_conv-tasnet_2019} or DPRNN \cite{luo_dual-path_2020}  see a drastic degradation in performance when increasing the number of output sources they consider.
As an alternative approach, we use the speaker embedding-based stitching method proposed by \cite{coria_overlap-aware_2021}.
Speaker embeddings can be used to solve the permutation problem between different local predictions, but also to distinguish between new and already seen speakers.

\begin{figure}[t]
    \centering
    \includegraphics[width=0.4\textwidth, height=3cm]{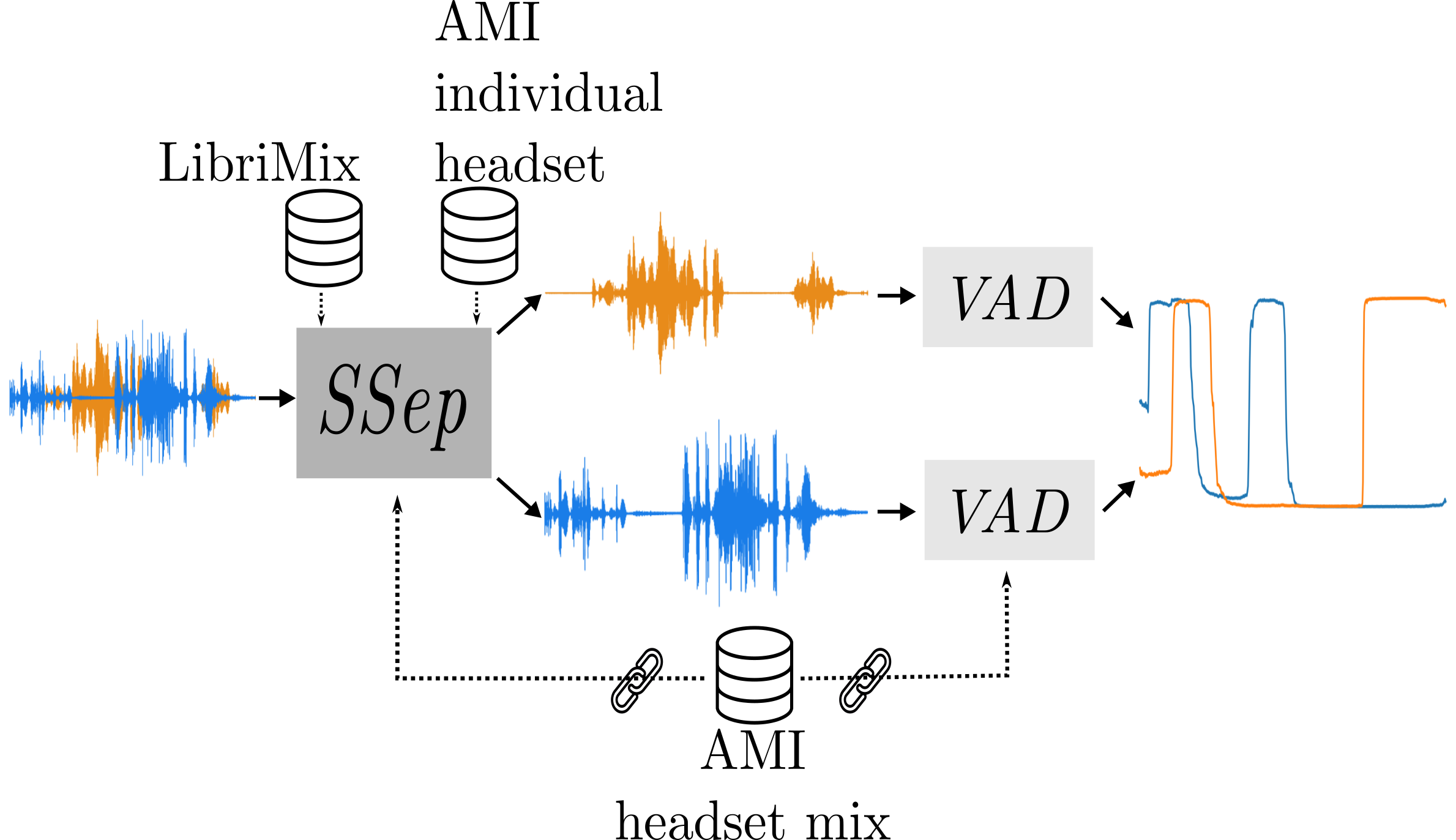}
    \caption{Diagram of the inference process for local predictions on 5-s windows.
    The dataset used for the end-to-end finetuning is symbolized with chains.
    }
    \label{figure_block}
\end{figure}

\section{Proposed system}
The system proposed in this work is composed of 3 components: speech separation (SSep), voice activity detection (VAD) and a speaker embedding-based stitching mechanism.\par
Speech separation is performed on sliding 5-s windows to obtain ``local'' predictions.
The overlap between subsequent windows is 90\%, meaning the step is 500 ms.
For each window, the active speakers for the incremental clustering are searched only in the last 500 ms, while the rest of the window is used as context to better estimate the speaker embeddings.

\vspace{3pt}
\noindent \textbf{Separation and VAD.} \
The 5-s input segments \(x \in \mathbb{R}^{1 \times T}\) are first fed to the SSep model, which estimates the sources \(\hat{s}_{i} \in \mathbb{R}^{1 \times T}\) for each output of the model, where \(T\) is the number of samples in the segments.
VAD is then applied independently to each \(\hat{s}_{i}\) to estimate the speech activities \(\hat{a}_{i} \in [0, 1]^{1 \times F}\), where F is the number of frames.
The SSep model takes a single-channel audio as input and outputs a fixed number of estimated sources.
In this work, we test models with 2 or 3 output sources.
To bridge the domain shift between real data and estimated sources, the VAD needs to be finetuned.
Similarly to \cite{morrone_end--end_2023}, we consider two types of finetuning.
The first variant is to adapt only the VAD on the estimates of the SSep model.
The second finetuning strategy consists of jointly adapting both the SSep and the VAD in an end-to-end fashion.

\vspace{3pt}
\noindent \textbf{Speaker embedding-based stitching.} \
To combine local predictions across time, a permutation problem needs to be solved between consecutive windows.
Additionally, new speakers can appear as well.
In this work, we rely on the use of speaker embeddings for the sake of stitching together the predictions on the 5-s sliding windows, following the approach from \cite{coria_overlap-aware_2021}.\par
\begin{figure}
    \centering
    \includegraphics[width=0.5\textwidth]{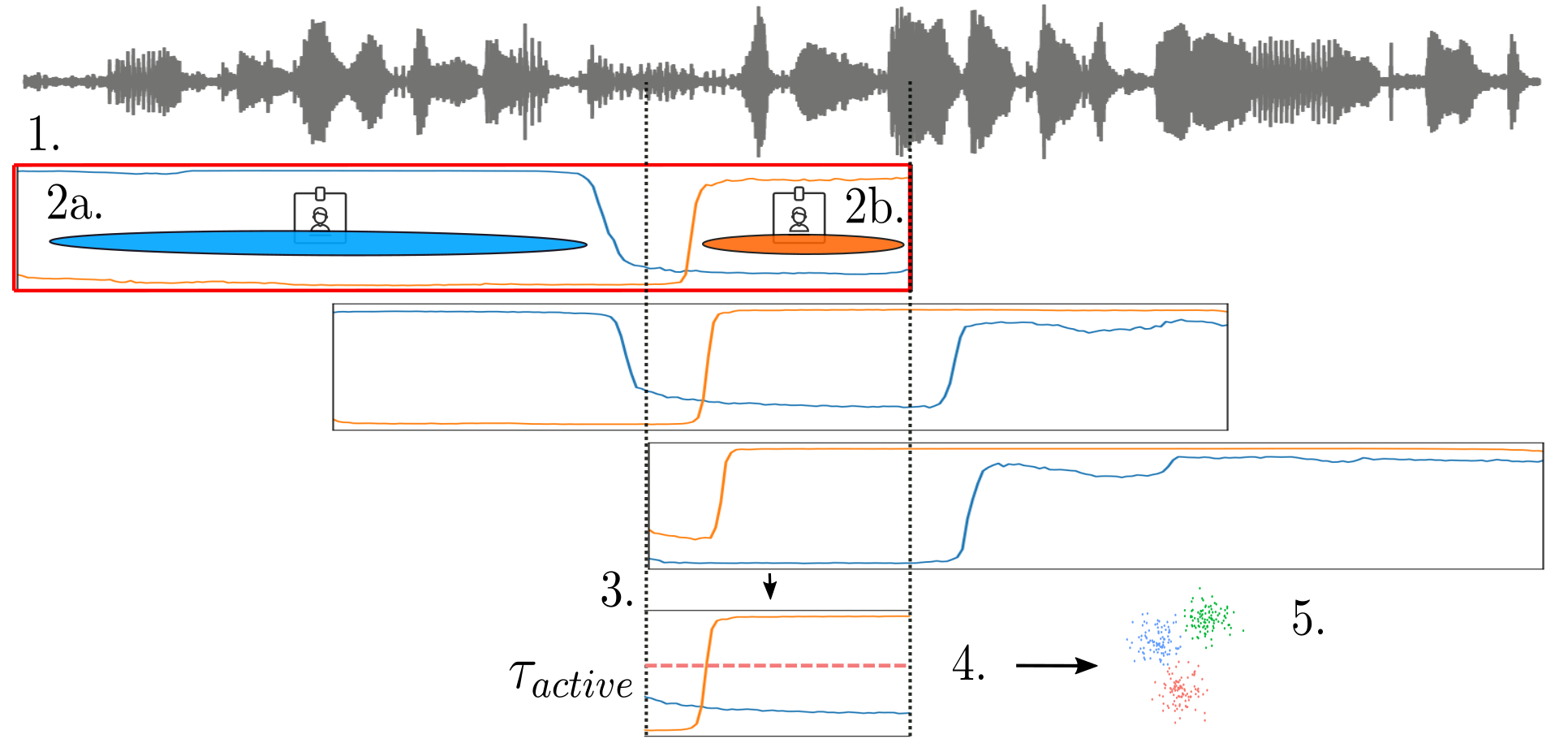}
    \caption{
    Diagram of a single step of the stitching of local predictions.
    }
    \label{figure_coria_system}
\end{figure}

Figure \ref{figure_coria_system} summarizes the logic of the stitching process.
At each step: 1) start from the current window (top, outline red), 2a/2b) speaker embeddings are estimated, 3) the predictions of the active speakers are aggregated with a delay (bottom) if the latency is above the minimum of 500ms, 4) the activities are binarized using \(\tau_{active}\) to get speaker segments, 5) incremental clustering is performed on the speaker embeddings to find the best match between the speaker segments and the existing centroids.
All windows in the figure are shown with the estimated activities already computed.
To improve the statistic pooling layer, the estimated activities are used to inform the weights of the frames when computing the embeddings, as detailed in \cite{coria_overlap-aware_2021}.
\par
The clustering is governed by other two hyperparameters: \(\delta_{new}\) and \(\rho_{update}\).
The first parameter, \(\delta_{new}\), defines the threshold distance between a new embedding and the closest centroid to define a new speaker.
The distance metric used is the cosine similarity, as the speaker embedding is an implementation of the X-vector architecture \cite{snyder_deep_2017, snyder_x-vectors_2018} trained with additive angular margin loss \cite{deng_arcface_2019}.
The latter parameter, \(\rho_{update}\), prevents embeddings estimated from short speech segments from updating the centroids of their cluster.
The rationale is to prevent noisy embeddings from damaging the speaker representation of the centroids.

\section{Experimental setup}
\noindent \textbf{Dataset.} \
As this work focuses on meeting conversations, we evaluate our models on the AMI dataset \cite{mccowan_ami_2005}. 
Specifically, the evaluation of all our proposed models is performed on the headset mix, our focus being on single-microphone settings.
The other datasets used in the training and finetuning stage of the SSep models are LibriMix type ``mix\_both" \cite{cosentino_librimix_2020} and the individual (speaker-focused) headset recordings from AMI.
In order to compare our results to previous works, we have used the AMI evaluation protocol proposed in \cite{landini_2022bayesian_2022}.

\vspace{3pt}
\noindent \textbf{Architecture configuration and training details.}\label{sec:training_details} \ We consider two different separation architectures: ConvTasNet \cite{luo_conv-tasnet_2019} and DPRNN \cite{luo_dual-path_2020}, with a view to gaining insight on the behaviour of our system, especially its robustness, when considering different SSep architectures.
Both separation models are first trained on fully overlapped mixtures from LibriMix type ``mix\_both"\cite{cosentino_librimix_2020}, using 3-second segments.
For both models, we have used the same configuration as in the Asteroid toolkit \cite{pariente_asteroid_2020}. However, for DPRNN, to reduce the computational burden, the kernel size and the stride have been set, respectively, to 32 and 16. 
The chunk size has been increased to 300 to reduce the length of the inter-RNN processing.
The hop size was increased to 150 to maintain it at 50\% of the chunk size.
\par
After the training on LibriMix, the SSep models are finetuned on real data using the AMI train set, lowering the learning rate by a factor of 10 to 0.0001.
Given that the isolated sources are not available, we have resorted to using the individual headset microphones of the active speakers as the ground truth.
Note that these recordings may not be optimal as sources because they include other speakers in the vicinity.
Finally, as a last finetuning step, we have joined the VAD to each output source of the SSep model.
In our experiments, we have used the pretrained VAD from Pyannote \cite{bredin_end--end_2021} \footnote{available at \href{https://huggingface.co/pyannote/embedding}{hf.co/pyannote/embedding/}. Note that this is not the same model as the one used in \cite{coria_overlap-aware_2021}, which exploits an improved version instead.}.
We have tried two combinations: freezing the SSep model to finetune only the VAD, and finetuning the entire system end-to-end.
While both approaches showed remarkable improvements over the use of a pre-trained VAD, the end-to-end approach has shown the best performance, as shown in Table \ref{table_sota_and_ablation} and discussed in Section \ref{sec:model_selection}


\section{Results}
All results presented rely on the AMI protocol presented in \cite{landini_2022bayesian_2022}.
The inference is carried out under full evaluation, meaning with no collar and evaluating also overlapped speech.

\begin{table}[ht]
\centering
\renewcommand{\arraystretch}{0.7} 
\begin{tabular}{lcccc}
\toprule
Model & DER & FA & MS & SC \\
\midrule
SSep AMI + VAD E2E &\textbf{27.2} & \textbf{1.8} & 18.4 & \textbf{7.0} \\
SSep LibriMix + VAD E2E & 28.4 & 2.0 & \textbf{18.3} & 8.1 \\
SSep LibriMix + VAD finetuned & 34.4 & 1.9 & 22.6 & 9.9 \\
SSep LibriMix + VAD & 42.8 & 3.7 & 19.3 & 19.8 \\
\specialrule{.05em}{.05em}{.05em}
Coria et al.\tablefootnote{Reproduced results} \cite{coria_overlap-aware_2021}  & 28.5 & 4.4 & \textbf{12.0} & 12.1 \\
\specialrule{.10em}{.05em}{.05em}
Kwon et al. \cite{kwon_absolute_2022} & 22.9 & n.a. & 14.5 & 8.3 \\
Yue et al. \cite{yue_online_2022} & 19.0 & - & - & - \\
Kynych et al. \cite{ekstein_online_2023} & 21.2 & - & - & - \\ 
\bottomrule
\end{tabular}
\caption{Comparison of our proposed online diarization system with the literature. The top section of the table presents an ablation study of the training methodology. The SSep used is ConvTasNet with 2 outputs. The bottom section of the table report results which rely at least in part on oracle information.}
\label{table_sota_and_ablation}
\end{table}

\vspace{3pt}
\noindent \textbf{Model performance and ablations.}\label{sec:model_selection} \
Table \ref{table_sota_and_ablation} presents the results of our new speaker diarization system based on SSep and VAD fine-tuned end-to-end, along with a few ablations allowing one to clarify the impact of each component of the system.
The results show that our proposed system is competitive with the previous work \cite{coria_overlap-aware_2021}, and that it improves the performance as measured by the overall DER. \par
Comparing the works from the bottom part of Table \ref{table_sota_and_ablation} with our methods is not straightforward as they use oracle information.
The only comparison that we can make is against VBx \cite{landini_2022bayesian_2022} (on which \cite{yue_online_2022} is based), but in an offline setting.
To this end, we use the speaker diarization pipeline from \cite{bredin_pyannoteaudio_2023}, which can be considered as an offline variant of \cite{coria_overlap-aware_2021}.
In this case, our best model achieves a DER of 23.5\%  without any hyperparameter tuning, while VBx with Pyannote VAD (instead of oracle VAD) achieves 24.1\% \cite{bredin_end--end_2021}.
\par
Further, the ablation experiments (upper part of Table \ref{table_sota_and_ablation}) show that all the components of our system play a role in improving performance.
Removing the AMI finetuning of the SSep model lowers performance even if the model is still finetuned end-to-end.
In line with \cite{morrone_end--end_2023}, switching from end-to-end finetuning to VAD-only adaptation degrades the performance.
Finally, a pre-trained SSep model with a non-finetuned VAD leads to the worst performance overall.

\vspace{3pt}
\noindent \textbf{Choice and parametrisation of the SSep model.} \
To test the dependence of the results on the quality of the SSep system used, we repeat the online inference on AMI with two different SSep models, with 2 or 3 output sources, for a total of four model combinations as presented in Table \ref{table_generalization_other_SSep}.
The results show that all the models considered are competitive with \cite{coria_overlap-aware_2021}.
For both ConvTasNet and DPRNN, the 2-output models achieve a better score than the 3-output counterpart.
On the other hand, the 2-output models obtain the worst missed speech result, which is expected as the distribution of 5-s segments with more than 2 speakers in AMI is non negligible.\par
\begin{figure}
    \centering
    \includegraphics[scale=0.20]{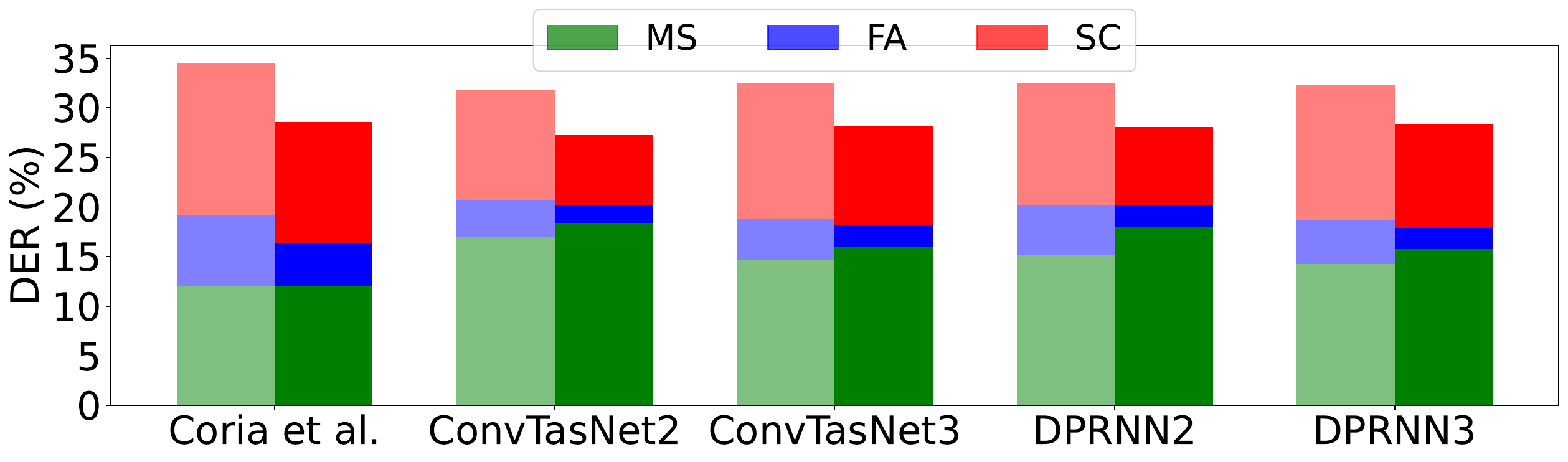}
    \caption{All proposed online systems compared to Coria et al.'s, tested at minimum (0.5s, left bar) and maximum latency (5s, right bar). The DER is broken down into its constituents: Missed Speech (MS), False Alarm (FA) and Speaker Confusion (SC).}
    \label{figure_latency_benchmark}
\end{figure}

The benchmark of the different SSep models is further explored in Figure \ref{figure_latency_benchmark}.
Relative to Coria et al \cite{coria_overlap-aware_2021}, our system improves the performance in low algorithmic latency settings.
For all SSep models considered, increasing the latency leads to a reduction of false alarms and speaker confusion, but also to an increase in missed detection. 
This is in contrast with \cite{coria_overlap-aware_2021}, for which missed speech seems not to be affected by the change in latency.

\begin{table}[ht]
\centering
\renewcommand{\arraystretch}{0.7} 
\begin{tabular}{lcccc}
\toprule
Model & DER & FA & MS & SC \\
\midrule
ConvTasNet2 & \textbf{27.2} & \textbf{1.8} & 18.4 & \textbf{7.0} \\
ConvTasNet3 & 28.1 & 2.1 & 16.0 & 10.0 \\
DPRNN2 & 28.0 & 2.2 & 18.0 & 7.8 \\
DPRNN3 & 28.4 & 2.1 & 15.8 & 10.4 \\
\bottomrule
\end{tabular}
\caption{Online diarization results for SSep AMI + VAD E2E using different SSep models with 2 or 3 output sources.}
\label{table_generalization_other_SSep}
\end{table}

\begin{table*}[t]
\centering
\renewcommand{\arraystretch}{0.7} 
\newcolumntype{Y}{>{\centering\arraybackslash}X}  
\begin{tabularx}{\textwidth}{@{} p{2.8cm} *{6}{Y} @{}}
\toprule
\multirow{2}{=}{Model} & \multicolumn{4}{c}{Test DER by number of speakers} & \multirow{2}{=}{Test DER} \\
                  \cmidrule(lr){2-5}
                  & 1 spk  & 2 spks & 3 spks & 4 spks & \\ \midrule
ConvTasNet2 & \text{6.0} $\pm$ \text{0.3} & \text{14.8} $\pm$ \text{0.4} & \text{25.3} $\pm$ \text{0.5} & \text{34.6} $\pm$ \text{0.8} & \text{15.8} $\pm$ \text{0.3} \\
\mbox{\quad \emph{OVL-only scoring}} & \emph{n.a.} & \emph{\text{16.7} $\pm$ \text{0.4}}
 & \emph{\text{26.7} $\pm$ \text{0.4}} & \emph{\text{33.0} $\pm$ \text{0.5}}
 & \emph{\text{24.3} $\pm$ \text{0.3}}
 \\
ConvTasNet3 & \text{5.9} $\pm$ \text{0.4} & \text{16.4} $\pm$ \text{3.8} & \text{23.9} $\pm$ \text{0.6} & \text{30.3} $\pm$ \text{0.1} & \text{15.4} $\pm$ \text{0.3} \\
\mbox{\quad \emph{OVL-only scoring}} & \emph{n.a.} & \emph{\text{21.1} $\pm$ \text{0.4}}
 & \emph{\text{24.2} $\pm$ \text{0.5}}
 & \emph{\text{28.7} $\pm$ \text{0.8}}
 & \underline{\emph{\text{24.0} $\pm$ \text{0.3}}}
 \\ 
DPRNN2 & \text{6.8} $\pm$ \text{0.3} & \text{15.7} $\pm$ \text{0.5} & \text{25.3} $\pm$ \text{0.5} & \text{35.3} $\pm$ \text{0.9} & \text{16.4} $\pm$ \text{0.3} \\
\mbox{\quad \emph{OVL-only scoring}} & \emph{n.a.} & \emph{\text{17.7} $\pm$ \text{0.4}}
 & \emph{\text{27.6} $\pm$ \text{0.5}}
 & \emph{\text{33.9} $\pm$ \text{0.6}}
 & \emph{\text{25.2} $\pm$ \text{0.3}}
 \\
DPRNN3 & \text{5.4} $\pm$ \text{0.3} & \text{15.0} $\pm$ \text{0.5} & \text{24.7} $\pm$ \text{0.4} & \text{33.1} $\pm$ \text{0.8} & \textbf{\text{15.2} $\pm$ \text{0.3}} \\
\mbox{\quad \emph{OVL-only scoring}} & \emph{n.a.}  & \emph{\text{18.8} $\pm$ \text{0.4}}
 & \emph{\text{25.6} $\pm$ \text{0.4}}
 & \emph{\text{32.1} $\pm$ \text{0.7}}
  &  \emph{\text{24.5} $\pm$ \text{0.3}}
 \\
\specialrule{.10em}{.05em}{.05em}
Coria et al. \cite{coria_overlap-aware_2021}  & \text{5.9} $\pm$ \text{0.3} & \text{17.0} $\pm$ \text{0.5} & \text{26.9} $\pm$ \text{0.6} & \text{33.5} $\pm$ \text{0.9}  & \text{16.7} $\pm$ \text{0.2} \\
    \mbox{\quad \emph{OVL-only scoring}}  & \emph{n.a.} & \emph{\text{24.1} $\pm$ \text{0.6}}
 &  \emph{\text{29.6} $\pm$ \text{0.5}}
 & \emph{\text{32.7} $\pm$ \text{0.6}}
  & \emph{\text{28.2} $\pm$ \text{0.3}}
  \\ 
\bottomrule
\end{tabularx}
\caption{Performance on individual segments of 5 seconds. The error on segments with no speakers is not reported because it is null for all the models. The performance scoring only the overlapped portion of the speech is noted as \emph{OVL-only scoring}. For all models \(\tau_{active}=0.5\). The results are reported with a 95\% confidence interval.}
\label{table_individual_segments}
\end{table*}

\noindent \textbf{Local inference and overlapped speech performance.}  To disentangle the contributions of the local prediction from those of the stitching mechanism, we have evaluated the SSep models on individual segments of 5 seconds, as detailed in Table \ref{table_individual_segments}.
The only hyperparameter here is the threshold value to convert the continuous prediction into binarized outputs, equivalent to \(\tau_{active}\).
For each model, we also report the performance when scoring only overlapped speech sections.
The baseline for comparison is the segmentation model from \cite{coria_overlap-aware_2021}, an LSTM-based EEND model with 4 outputs trained on multiple datasets including AMI. \par
The results show that all our proposed models improve on the baseline, both regarding overall test DER and considering only overlapped speech.
Interestingly, in contrast with the results in Table \ref{table_generalization_other_SSep}, here we find DPRNN to perform better that ConvTasNet.
Also, 3-output models perform better than the 2-output ones.
It is important to note that all systems are competitive also on segments with only one speaker, which can be mishandled by SSep models, as discussed in \cite{fang_deep_2021, niu_separation_2021}.
Additionally, SSep models with 2 and 3 outputs have similar performance on segments with 1 and 2 speakers, which is not to be expected if one considers the results from Section \ref{sec:SS_different_outputs}.
We attribute this generalization to the end-to-end finetuning, as models are trained also on segments with fewer speakers, contrary to training with SI-SDR loss. 
With these results, we claim that the SSGD framework can be robust enough to be used as a stand-alone approach, without being integrated with other methods like in \cite{fang_deep_2021, niu_separation_2021}.

\vspace{3pt}
\noindent \textbf{Behaviour of SSep models after adaptation on real data.} \
The SSGD framework is appealing also because it performs separation for free.
For each step of the training pipeline detailed in Section \ref{sec:training_details}, we show some examples of how the SSep models behave\footnote{\href{https://github.com/egruttadauria98/SSpaVAlDo}{egruttadauria98/SSpaVAlDo}}.
It is not possible to objectively evaluate the separation on AMI because ground-truth sources are not available.
Here we limit ourselves to a few observations on the behaviour of the models.\par
The SSep models are first pretrained on fully overlapped mixtures from LibriMix, with as many speakers as the number of outputs of the model.
For a SSep model trained on fully overlapped mixtures, all recordings with less speakers than the number of outputs are out-of-domain examples.
After finetuning the SSep models on real data from AMI, the estimated sources were found to be less affected by phenomena that lead to speaker confusion, such as splitting one speaker into multiple outputs.
Nevertheless, because the SSep models are finetuned on individual microphones which contain also speech from nearby speakers, the estimated sources present more leakage than the models just trained on LibriMix.
The leaked speakers are always at lower energy than the main speaker, so a finetuned VAD is usually able to distinguish them and avoid false alarms.
With the end-to-end finetuning, the SSep models learn to make a few little adjustments to improve the diarization score, but the leakage is still present.
As such, finetuning end-to-end alone does not lead to better separation automatically, as the goal is only diarization performance.
Our interpretation is that the finetuning end-to-end pushes the model to reduce the leakage at least when it can lead to false alarm, while it is otherwise kept. 

\begin{figure}
    \centering
    \includegraphics[scale=0.20]{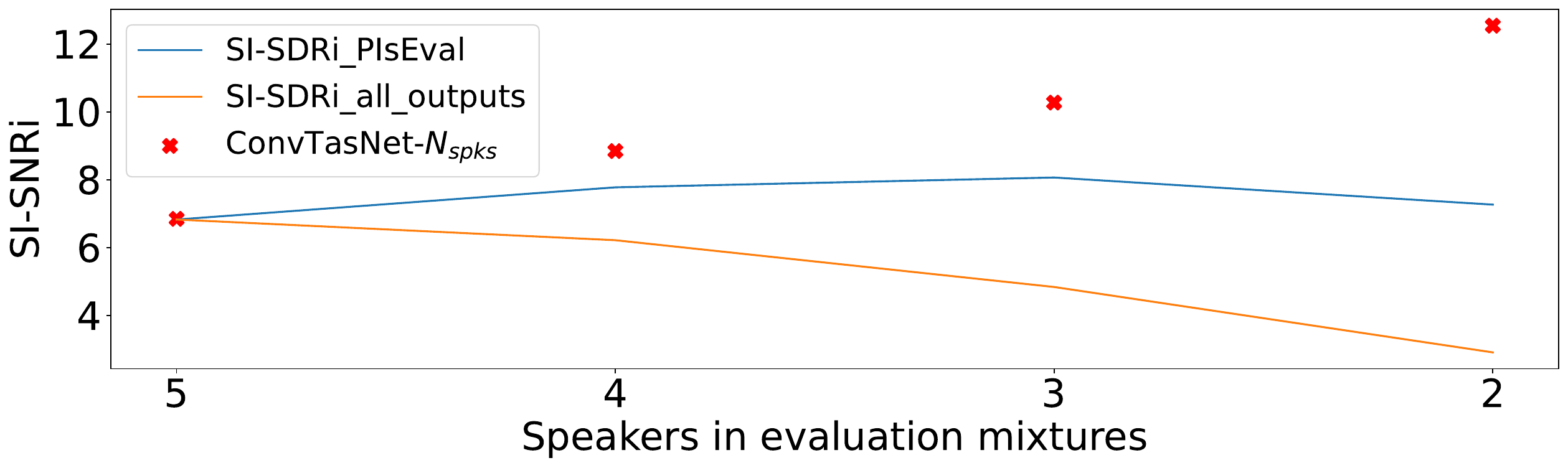}
    \caption{Performance of ConvTasNet5 trained on Libri5Mix on mixtures with 5, 4, 3 and 2 speakers. The red crosses show the performance of the SSep model with as many outputs as the speakers in the mixtures.
    }
    \label{figure_ss_models_outputs}
\end{figure}

\vspace{3pt}
\noindent \textbf{Relationship between SSep performance and number of outputs.}\label{sec:SS_different_outputs} \
We have found that increasing the number of outputs of the speech separation model always leads to a loss in performance.
Figure \ref{figure_ss_models_outputs} shows how the performance of ConvTasNet5  changes when testing it on mixtures with 5 or fewer speakers, as shown on the x-axis.
As there is no straightforward way to evaluate a SSep model with a mismatch between the estimated outputs and the actual number of sources, we use both a harsh metric and a forgiving metric.
The harsh metric, \emph{all outputs}, used as a reference for the additional sources an all zero-signal \footnote{A small constant is added to avoid numerical errors in the computation of the SI-SDR metric}.
The forgiving metric, \emph{PIsEval}, uses the oracle number of speakers in the mixture, \(N_{spks}\), to score only the estimated sources that best resemble the references.
For ConvTasNet5, \emph{PIsEval} improves initially when \(N_{spks}\) is reduced first to 4 and then to 3, because the mixtures are easier to separate.
Once the \(N_{spks}\) reaches 2, the performance worsens, possibly because the out-of-domain factor outweighs the easier separation.
Lastly, we also plot with red crosses the performance of ConvTasNet-\(N_{spks}\) on mixtures with \(N_{spks}\) speakers.
At each value of the x-axis, the difference between the red cross and the blue line shows the minimum loss in performance by using a SSep with 5 outputs instead of a SSep with as many outputs as the speakers in the mixtures.

\section{Conclusions}
We have presented a novel SSGD system for online speaker diarization that achives state-of-the-art performance on AMI headset mix.
Our results show that the limitations of SSep on real data can be overcome, leading to a diarization model that can better handle overlapped speech and estimates sources for each speaker.

\bibliographystyle{IEEEtran}
\bibliography{sample}

\end{document}